\documentclass[a4paper,conference]{IEEEtran}

\usepackage{indentfirst}
\usepackage[T1]{fontenc}
\usepackage[]{graphicx}
\usepackage[]{epsfig}
\usepackage{amsmath}
\usepackage{amssymb}
\usepackage{verbatim}
\usepackage{tikz} 
\usepackage{bbm}
\usepackage{enumerate}
\usepackage{pdfsync}
\usepackage{algorithm}
\usepackage{algpseudocode}
\usepackage{array}
\usepackage{mathrsfs}
\usepackage{dsfont}

\usetikzlibrary{calc} 
\usetikzlibrary{snakes,arrows} 

\usepackage[linktoc=page,colorlinks=true,linkcolor=titres,citecolor=titres]{hyperref}
\usepackage{bookmark}
\usepackage{makeidx}
\usepackage{amsfonts}
\usepackage{amsthm}
\usepackage{url}
\usepackage[caption=false,font=footnotesize]{subfig}
\usepackage{cite}
\usepackage{color}
\usepackage{tocvsec2}

\newtheorem{theorem}{Theorem}
\newtheorem{definition}{Definition}

\newcommand{\E}[1]{\mathbf{E}_{}\!\left[#1\right]}

\definecolor{int}{rgb}{0.8,0.8,1}
\definecolor{ext}{rgb}{0,0,1}
\definecolor{redbox}{RGB}{191,18,56} %
\definecolor{blackbox}{RGB}{0,0,0} %
\definecolor{brownbox}{RGB}{128,99,90} %

\definecolor{clair}{rgb}{0.910,0.933,0.957}
\definecolor{moyen}{rgb}{0.102,0.349,0.557}
\definecolor{fonce}{rgb}{0.067,0.231,0.369}
\definecolor{titres}{rgb}{0.137,0.466,0.741}
\definecolor{apricot}{rgb}{0.961,0.506,0.216}
\definecolor{grey}{rgb}{0.58,0.58,0.58}

\begin{document}

\pgfdeclarelayer{background layer} 
\pgfdeclarelayer{foreground layer} 
\pgfsetlayers{background layer,main,foreground layer}

\title{Building a coverage hole-free communication tree}
\author{\IEEEauthorblockN{Ana\"is Vergne,
Laurent Decreusefond, and
Philippe Martins}\\
\IEEEauthorblockA{LTCI, T\'el\'ecom ParisTech, Universit\'e Paris-Saclay, 75013, Paris, France}
}

\maketitle
\begin{abstract}
Wireless networks are present everywhere but their management can be
tricky since their coverage may contain holes even if the network
is fully connected. In this paper we propose an algorithm that can
build a communication tree between nodes of a wireless network with
guarantee that there is no coverage hole in the tree. We use
simplicial homology to compute mathematically the coverage, and Prim's
algorithm principle to build the communication tree. Some simulation
results are given to study the performance of the algorithm and
compare different metrics. In the end, we show that our algorithm can
be used to create coverage hole-free communication groups with a limited
number of hops.
\end{abstract}

\section{Introduction}
Wireless networks are everyday more present in our lives: WiFi is the
main internet access in our homes, cellular systems such as 4G and
soon 5G provide its access everywhere else. Moreover with IoT, every
object in our kitchen or in our bathroom will in the near future be
connected as well. When managing a network, it is often useful to
build a communication tree of the network nodes in order to transmit
messages to every node efficiently. The spanning tree is the
answer to that problem: the fact that it is a tree guarantees that
there are no superficial links, and spanning means that all connected
nodes are included. Several well-known algorithms allow to find the
minimum spanning tree in a graph according to a given metric. We can
cite Kruskal's algorithm \cite{kruskal_shortest_1956}, Prim's
algorithm \cite{prim_shortest_1957}, or Bor\r{u}vka's algorithm
\cite{boruvka_1926}. 

However, the quality of service of wireless network is
primarily providing access to its users, in other terms provide
coverage. Therefore, a communication tree with coverage holes could be
pointless. Meanwhile, deciding whether a set of base stations does cover
a whole domain is not that easy when the network is irregularly
deployed, as it is the case for cellular networks see
\cite{deng_ginibre_2015} or \cite{gomez_case_2015}. Based on the
geometrical data of the network, we can build a combinatorial object
to represent it: the simplicial complex. Basically a simplicial
complex is the generalization of the concept of graph, it is made of
$k$-simplices where $0$-simplices are vertices, $1$-simplices are
edges, $2$-simplices are triangles, $3$-simplices are tetrahedron and
so on. In particular, geometrical simplicial complexes such as the
\u{C}ech complex and the Vietoris-Rips complex allows to represent
exactly and approximately the coverage of the union of the coverage
disks as stated in the Nerve lemma
in~\cite{ghrist_coverage_2005}. Then algebraic topology,
\cite{hatcher_algebraic_2002}, is the mathematical tool used to
compute the number of connected components, of coverage holes, and of
3D voids, that are the so-called Betti numbers of the simplicial
complex representing the network, as detailed in
\cite{de_silva_coordinate-free_2006}.  

In this article, we introduce an algorithm that can
build a communication tree between the connected nodes of a wireless
network with guarantee that there is no coverage hole in the
tree. First, we use simplicial homology to represent the network, and
algebraic topology to compute its coverage. Then we modify Prim's
algorithm in order to only select vertices that do not create coverage
holes. We provide simulation results to measure the performance of our
algorithm in terms of number of rejected nodes, and surface of covered
area. We then compare different metrics for the weight of edges, and
find that the height metric, from the simplicial complex
representation, provides results with the shortest branches both in
terms of hops and length without losing any covered area. Finally, we
extend our algorithm to build coverage hole-free communication trees
in larger networks.

This is the first algorithm of this type that we know of. Finding a
spanning tree in a graph is an old and classic problem
\cite{kruskal_shortest_1956, prim_shortest_1957, boruvka_1926}. But
the use of simplicial homology for wireless networks is just about a
decade old \cite{ghrist_coverage_2005}. Since, the computational time
to obtain the Betti numbers can explode with the size of the
simplicial complex, many works focus on faster ways to compute them,
for instance in a decentralized way \cite{muhammad_decentralized_2007}, 
using persistent homology \cite{zomorodian_computing_2005}, thanks to
chain complexes reduction \cite{kaczynski_homology_1998} , or with
witness complexes reduction \cite{de_silva_topological_2004}.
Simplicial complexes reduction can also be used for coverage hole
detection \cite{yan_homology-based_2015} and energy efficiency
in cellular networks \cite{vergne_simplicial_2015}.

In Section \ref{sec_cov}, we provide the mathematical background. Then
in Section \ref{sec_algo}, we give the algorithm for building a
coverage hole-free communication tree along with
simulation results in Section \ref{sec_sim}. Finally we propose an extension to the building of
communication groups in Section \ref{sec_forest}.

\section{Coverage of a network}
\label{sec_cov}
\subsection{Simplicial homology and algebraic topology}
Considering a set of points representing network nodes, the first idea
to apprehend the topology of the network would be to look at the
neighbors graph: if the distance between two points is less than a
given parameter then an edge is drawn between them. However this
representation is too limited to transpose the network's
topology. First, only $2$-by-$2$ relationships are represented in the
graph, there is no way to grasp interactions between three or more
nodes. Moreover, there is no concept of coverage in a graph. That is
why we are interested in more complex objects.

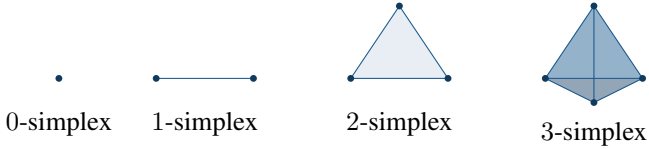
\begin{figure}[h]
  \centering
    \begin{tikzpicture}[scale=0.64]
\fill [color=fonce] (0,0) circle (2pt);
\node [below] at (0,-0.5) {$0$-simplex};
\draw [color=moyen] (2,0)--(4,0);
\fill [color=fonce] (2,0) circle (2pt);
\fill [color=fonce] (4,0) circle (2pt);
\node [below] at (3,-0.5) {$1$-simplex};
\fill [color=clair] (6,0)--(8,0)--(7,1.5);
\draw [color=moyen] (6,0)--(8,0)--(7,1.5)--(6,0);
\fill [color=fonce] (6,0) circle (2pt);
\fill [color=fonce] (8,0) circle (2pt);
\fill [color=fonce] (7,1.5) circle (2pt);
\node [below] at (7,-0.5) {$2$-simplex};
\fill [color=moyen, opacity=1] (10,0)--(12,0)--(11,1.5);
\fill [color=fonce,opacity=1] (10,0)--(12,0)--(11,-0.5);
\fill [color=clair,opacity=0.6] (10,0)--(11,-0.5)--(11,1.5);
\fill [color=clair,opacity=0.6] (11,-0.5)--(12,0)--(11,1.5);
\draw [color=moyen] (10,0)--(12,0)--(11,1.5)--(10,0);
\draw [color=moyen] (10,0)--(11,-0.5)--(11,1.5);
\draw [color=moyen] (12,0)--(11,-0.5);
\fill [color=fonce] (10,0) circle (2pt);
\fill [color=fonce] (12,0) circle (2pt);
\fill [color=fonce] (11,1.5) circle (2pt);
\fill [color=fonce] (11,-0.5) circle (2pt);
\node [below] at (11,-0.7) {$3$-simplex};
    \end{tikzpicture}
  \caption{Examples of $k$-simplices.}\label{fig_simplices}
\end{figure}

Indeed, graphs can be generalized to more generic combinatorial
objects known as simplicial complexes. While graphs model binary
relations, simplicial complexes can represent higher order
relations. A simplicial complex is thus a combinatorial object made up
of vertices, edges, triangles, tetrahedra, and their $n$-dimensional
counterparts. 
Given a set of vertices $X$ and an integer $k$, a $k$-simplex is an
unordered subset of $k+1$ vertices $\{x_0,\dots, x_k\}$ where $x_i\in
X, \forall i \in \{0,\dots,k\}$ and $x_i\not=x_j$ for all
$i\not=j$. Thus, a $0$-simplex is a vertex, a $1$-simplex an edge, a
$2$-simplex a triangle, a $3$-simplex a tetrahedron, etc. See
Fig.~\ref{fig_simplices} for instance. 

\vspace{-0.2cm}

Any subset of vertices included in the set of the $k+1$ vertices of a
$k$-simplex is a face of this $k$-simplex. A $k$-face is then a face
that is a $k$-simplex. The inverse notion of face is coface. An
abstract simplicial complex is a set of simplices 
such that all faces of these simplices are also in the set of
simplices.

In this article, we are interested in representing the topology of a
wireless network, we introduce the following abstract simplicial complex:
\begin{definition}[Vietoris-Rips complex]
 Let $(X,d)$ be a metric space, $\omega$ a finite set of points in
  $X$, and $r$ a real positive number. The Vietoris-Rips
  complex of parameter $r$ of $\omega$, 
  $\mathcal{R}_{r}(\omega)$, is the abstract simplicial complex
  whose $k$-simplices correspond to the unordered $(k+1)$-tuples of
  vertices in $\omega$ which are pairwise within distance less than
  $r$ of each other.
\end{definition}

The Vietoris-Rips complex is easy to build 
since it is only based on the neighbors graph information.
Moreover it provides an approximation of the exact topology of the network,
which is given by the \u{C}ech complex (see the Nerve lemma in
\cite{ghrist_coverage_2005}). This approximation is quite good:
in the case of a random uncorrelated deployment with
network nodes deployed according to a Poisson point process the error
is less than $0.06\%$ in the computation of the covered area
\cite{yan_accuracy_2012}. An example of a Vietoris-Rips complex
representing a wireless network can be seen in Fig.~\ref{fig_net}.

Given an abstract simplicial complex, one can define an orientation on
the simplices by defining an order on the vertices, where a change in the
orientation, that is a swap between two vertices, corresponds to a
change in the sign. 
Then let us define the vector spaces of the $k$-simplices of a
simplicial complex, and the associated boundary maps:
\begin{definition}
  Let $S$ be an abstract simplicial complex.

 For any integer $k$,
  $\mathscr{C}_k(S)$ is the vector space spanned by the set of oriented
  $k$-simplices of $S$.
\end{definition}

\begin{definition}
Let $S$ be an abstract simplicial complex and $\mathscr{C}_k(S)$ the
vector space of its $k$-simplices for any $k$ integer.

The boundary map $\partial_k$ is defined as the linear
  transformation $\partial_k: \mathscr{C}_k(S)\rightarrow
  \mathscr{C}_{k-1}(S)$ which acts on the 
  basis elements $[x_0,\dots,x_k]$ of $\mathscr{C}_k(S)$ via:
\vspace{-0.1cm}
  \begin{eqnarray*}
    \partial_k [x_0,\dots,x_k]= \sum_{i=0}^{k}{(-1)}^i [x_0,\dots,x_{i-1},x_{i+1},\dots,x_k].
  \end{eqnarray*}
\end{definition}

For example, for a $2$-simplex we have:
\vspace{-0.3cm}
\begin{figure}[h]
\centering 
\begin{tikzpicture}[scale=0.7]
\coordinate (x1) at (0,0);
\coordinate (x2) at (1,1);
\coordinate (x3) at (2,0);
\coordinate (x4) at (5,0);
\coordinate (x5) at (6,1);
\coordinate (x6) at (7,0);
 \fill [color=clair] (x1)--(x2)--(x3);
\draw[draw=moyen] (x1)--(x2);
\draw[draw=moyen](x2)--(x3);
\draw[draw=moyen](x3)--(x1);
\draw[draw=moyen, arrows={-triangle 45}] (x4)->(x5);
\draw[draw=moyen, arrows={-triangle 45}] (x5)->(x6);
\draw[draw=moyen, arrows={-triangle 45}] (x6)->(x4);
\draw[->][color=fonce] (1,.2) arc (270:-30:.2);
\draw [fill=fonce] (x1) circle (2pt);
\draw [fill=fonce] (x2) circle (2pt);
\draw [fill=fonce] (x3) circle (2pt);
\draw [fill=fonce] (x4) circle (2pt);
\draw [fill=fonce] (x5) circle (2pt);
\draw [fill=fonce] (x6) circle (2pt);
\node [below] at (x1) {$x_0$};
\node [above] at (x2) {$x_1$};
\node [below] at (x3) {$x_2$};
\node [below] at (x4) {$x_0$};
\node [above] at (x5) {$x_1$};
\node [below] at (x6) {$x_2$};
\node [below] at (1,-0.5) {\small $\partial_2([x_0,x_1,x_2])$};
\node [below] at (3,-0.75) {\small $=$};
\node [below] at (6,-0.5) {\small $[x_1,x_2]-[x_0,x_2]+[x_0,x_1]$};
 \end{tikzpicture}
\end{figure}

As its name indicates, the boundary map applied to a linear
combination of simplices gives its boundary. The boundary of a
boundary is the null application. Therefore the following theorem can be easily
demonstrated (see~\cite{hatcher_algebraic_2002} for instance):
\begin{theorem}
  For any $k$ integer, 
$\partial_k \circ\partial_{k+1}=0.$
\end{theorem}

Let $S$ be an abstract simplicial complex. Then we can denote 
the $k$-th boundary group of $S$ as $B_k(S)=\mathrm{im}
  \, \partial_{k+1}$, and the $k$-th cycle group of $S$ as
  $Z_k(S)=\ker \partial_{k}$. We have $B_k(S)\subset Z_k(S)$.
We are now able to define the $k$-th homology group and its dimension:
\begin{definition}
  The $k$-th homology group of an abstract simplicial complex $S$ is
  the quotient vector space: 
\vspace{-0.1cm}
  \begin{eqnarray*}
    H_k(S)=\frac{Z_k(S)}{B_k(S)}.
  \end{eqnarray*}
 The $k$-th Betti number of the abstract simplicial complex $S$ is: 
\vspace{-0.3cm}
  \begin{eqnarray*}
    \beta_k(S)=\dim H_k(S).
  \end{eqnarray*}
\end{definition}

According to its definition, the $k$-th Betti number counts the number
of cycles of $k$-simplices that are not boundaries of
$(k+1)$-simplices, that are the $k$-th dimensional holes. In small
dimensions, they have a geometrical interpretation:
\begin{itemize}
\item $\beta_0$ is the number of connected components,
\item $\beta_1$ is the number of coverage holes,
\item $\beta_2$ is the number of $3$D-voids.
\end{itemize}
For any $k\geq d$ where $d$ is the dimension, we have $\beta_k=0$.

We can now define the Euler characteristic of an abstract simplicial
complex:
\begin{definition}
  The Euler characteristic an abstract simplicial complex $S$ is the
  alternated sum of its Betti numbers:
\vspace{-0.1cm}
  \begin{eqnarray*}
   \chi(S)=\sum_{k=0}^{d-1}(-1)^k\beta_k.
  \end{eqnarray*}
But it can also be defined as:
\vspace{-0.1cm}
 \begin{eqnarray*}
   \chi(S)=\sum_{k=0}^{\infty}(-1)^ks_k,
  \end{eqnarray*}
where $s_k$ is the number of $k$-simplices in $S$.
\end{definition}

For further reading on algebraic topology, see
\cite{hatcher_algebraic_2002}.

\subsection{Percolation and coverage holes}
We are now considering coverage in light of percolation. Indeed when a
network is regularly deployed, think about the hexagonal model for
instance, if the network is connected then there is no coverage
hole. However, in real-life deployments, network cells are not
hexagons. When considering all the frequency bands owned by an
operator, network nodes are actually more similar to a Poisson point
process \cite{gomez_case_2015}. In this case, percolation does not
guarantee coverage.

In \cite{decreusefond_simplicial_2014}, the authors studied the
moments of the number of $k$-simplices for a
Vietoris-Rips complex based on a set of points drawn according to a
Poisson point process with the uniform norm on the $d$-dimensional
torus.
We are especially interested in the mean of the Euler characteristic:
 \begin{eqnarray*}
   \E{\chi(S)}=-\left(\frac{a}{r}\right)^d\sum_{k=0}^\infty
   \frac{(-\lambda r^d)^kk^d}{k!},
  \end{eqnarray*}
where $a$ is the side of the torus, $r$ is the Vietoris-Rips distance
for which two points are in the same simplex, $d$ is the dimension,
and $\lambda$ is the intensity of the Poisson point process.

\begin{figure}[h]
  \centering
    \includegraphics[width=5cm]{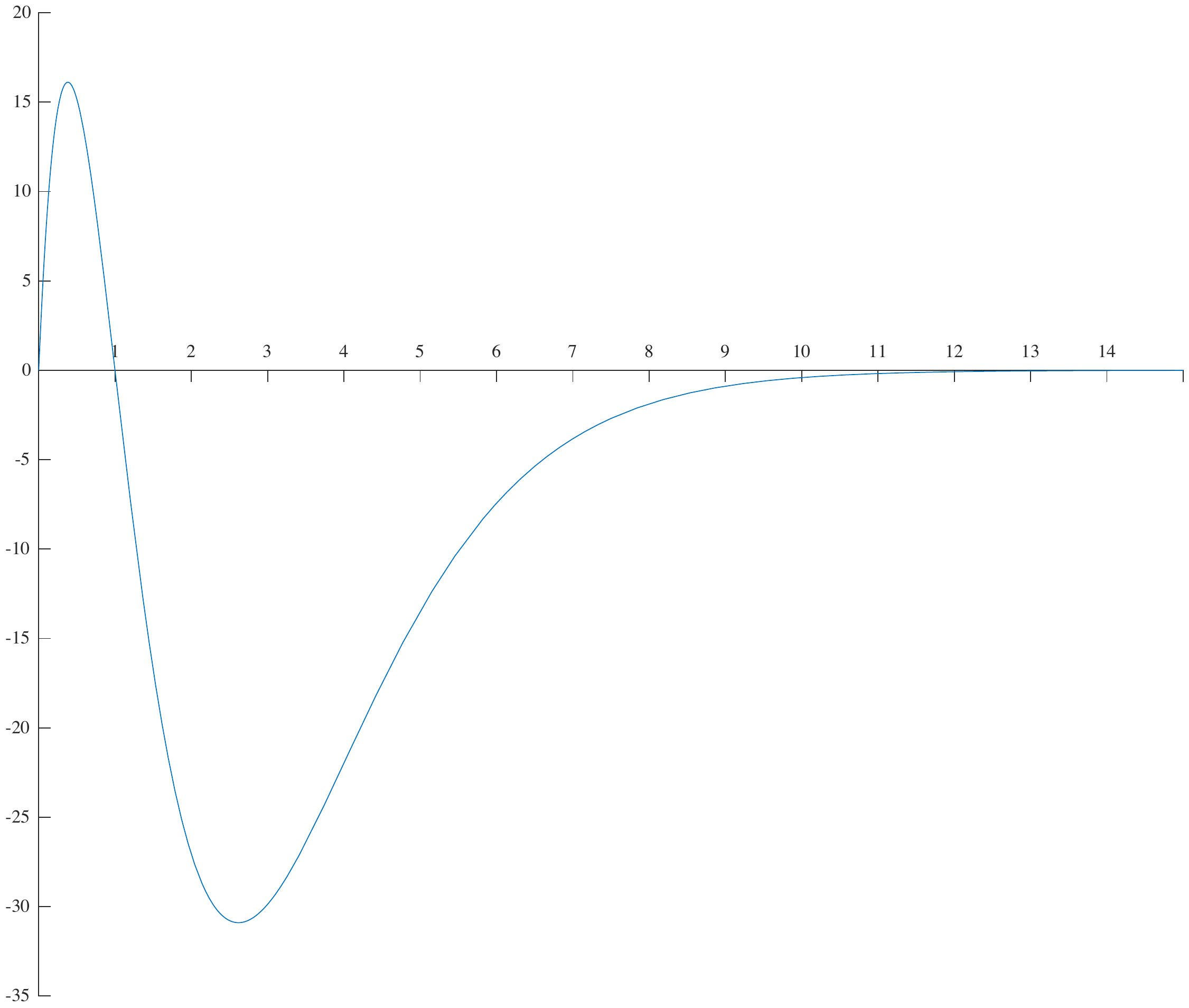}
\vspace{-0.1cm}
  \caption{ $\E{\chi(S)}$ for $\lambda = 0 ... 15$.}
\label{fig_chi}
\end{figure}

In two dimensions, this formula can be simplified:
 \begin{eqnarray*}
   \E{\chi(S)}=a^2\lambda (1-\lambda r^2) e^{-\lambda r^2}.
  \end{eqnarray*}
We plot $\E{\chi(S)}$ in function of $\lambda$ 
for $a\!=\!\!10$ and $r\!=\!\!1$
in Fig.~\ref{fig_chi}.

\vspace{-0.1cm}

However, in two dimensions, $\chi=\beta_0 - \beta_1$. Since the Betti
numbers are positive, we can interpret
the previous plot:
\begin{itemize}
\item When $\lambda$ is smaller than $0.5$, there are multiple
  connected components of just some points each, that are not large
  enough to create coverage holes. Then $\beta_0$ grows with $\lambda$
  and $\beta_1$
  is close to $0$. 
\item Around $\lambda = 0.5$, $\chi$ attains a maximum: percolation
  occurs. The number of connected components $\beta_0$ starts
  decreasing. On the other hand, coverage holes appear: $\beta_1$
  begins to increase.  
\item When $\lambda = 1$, $\chi$ becomes negative, that means that the
  number of coverage holes $\beta_1$ outnumbers the number of
  connected components $\beta_0$. $\beta_1$ goes on increasing
  while $\beta_0$ continues decreasing.
\item When $\lambda$ is greater than $3$, percolation has occurred:
  there is enough points to have only one connected component, and
  new points begin to fill coverage holes. That is to say that
  $\beta_0$ is close to $1$ and $\beta_1$ decreases. 
\item Finally when $\lambda$ is large enough,
  there is one unique component and no coverage hole: $\beta_0=1$ and
  $\beta_1=0$. 
\end{itemize}

From this, we can see that when network nodes are deployed randomly
following a Poisson point process, percolation occurs before full
coverage happens, and the network stays in this regime for many values of
$\lambda$. That means that while the network is connected and every
node can communicate with each other through a path of nodes, there
still exists regions that are uncovered. Therefore when simply building
a spanning tree, one is not sure not to include some coverage holes. That is
why we propose an algorithm for the construction of a coverage
hole-free communication tree.

\section{Coverage hole-free tree}
\label{sec_algo}
\subsection{Principle}
A spanning tree in a connected graph with $n$ vertices is a connected
subgraph of it which includes all of the $n$ vertices and has
exactly $n-1$ edges.
Finding a minimum or maximum spanning tree in a graph is a well-known
problem in computer science that is resolved by well-known algorithms
such as Kruskal's algorithm, Bor\r{u}vka's algorithm, and Prim's
algorithm. The minimum or maximum property is based on a weight associated with each
edge. It is possible to use any interesting metric: minimum distance,
maximum distance, or maximum redundancy for instance.

We are especially interested in Prim's algorithm since in this greedy
algorithm the spanning tree grows one edge at a time while staying
always connected \cite{prim_shortest_1957}. Indeed, at the beginning
of the algorithm, the tree is reduced to one vertex, chosen randomly.
Then at each step, the minimum-weight (or maximum-weight) edge  among
all the edges that join a vertex of the tree to a vertex outside the
tree is added to the tree. The algorithm stops when all vertices are
in the tree.

To build a coverage hole-free tree, our idea is simply to
modify the Prim's algorithm in order to check coverage at each step
thanks to simplicial homology, and to reject the edge, and consequently
its extremity vertex, if a coverage hole is created. Therefore, at the
end, a tree free of coverage holes is obtained.

\subsection{Algorithm}
First, our algorithm
computes the Vietoris-Rips complex based on the set of vertices and the
distance parameter given in input. It is important to note that we
only need to compute the complex up to the $2$-simplices since we are
only interested in the computation of $\beta_0$ and $\beta_1$. Then
the weights of the edges are computed according to a given metric.

After that, the tree $T$ is created with only the root, which is uniformly
drawn, and no edge. A set of 
potential edges $E_{\text{test}}$ with one extremity in the tree $T$ and the other
outside $T$ is defined. Then, while there are vertices outside
the tree and there are potential edges left, a potential edge of minimum
weight is added if it does not create a coverage hole. If it does, the
edge is removed from the set of potential edges.
We give in Algorithm~\ref{alg_prim} the pseudo-code. 

\begin{algorithm}[H]
  \caption{Coverage hole-free tree building algorithm.}
\label{alg_prim}
  \begin{algorithmic}[h]
    \Require{set $V$ of $n$ vertices, connection distance $r$.}
\State{Computation of 
  $S=\mathcal{R}_{r}(V)$\;}
\State{$E:=$ $\{1$-simplices of $S\}$   \textit{\%Edges of $S$}\;}
\State{Computation of the weights $\{w(e),e\in E\}$;}

\State{Draw uniformly a vertex $r \in V$ to be the root\;}
\State{$T:=\{r\}$  \textit{\%Tree}\;}
\State{$E_T:=\emptyset$   \textit{\%Edges of the tree}\;}
\State{$E_{\text{test}}:=(T,V\setminus T)$   \textit{\%Potential edges}\;}

\While{$|T|<n$ and $|E_{\text{test}}|>0$}
\State{Take $e$ such that
  $w(e)\!=\!\min_{}\{w(f), f\in E_{\text{test}}\}$\;}
\State{Let $x$ be the extremity of $e$ in $V \setminus T$\;}
\State{Computation of 
  $S_T=\mathcal{R}_{r}(T\cup\{x\})$ and $\beta_{1}(S_T)$\;}

\If{$\beta_{1}(S_T)\neq 0$}
\State{$E_{\text{test}}=E_{\text{test}}\setminus \{e\}$\;}
\Else{}
\State{$T:=T\cup\{x\}$\;}
\State{$E_T:=E_T\cup\{e\}$\;}
\EndIf{}

\EndWhile{}

\Return{$T$, $E_T$}
  \end{algorithmic}
\end{algorithm}

\begin{figure}[h]
  \centering
    \includegraphics[width=4.35cm]{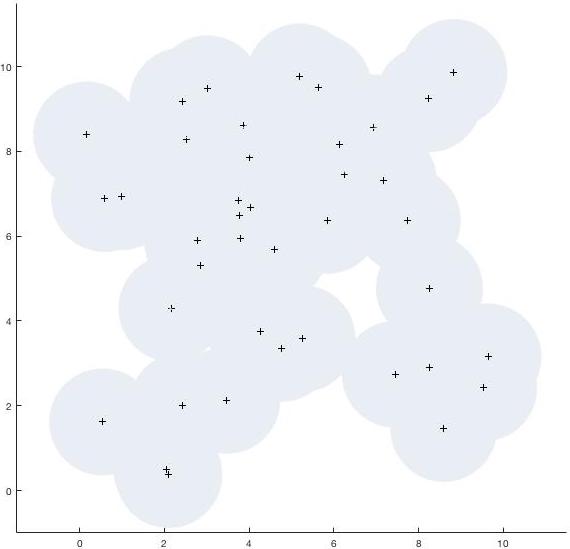}
 \includegraphics[width=4.35cm]{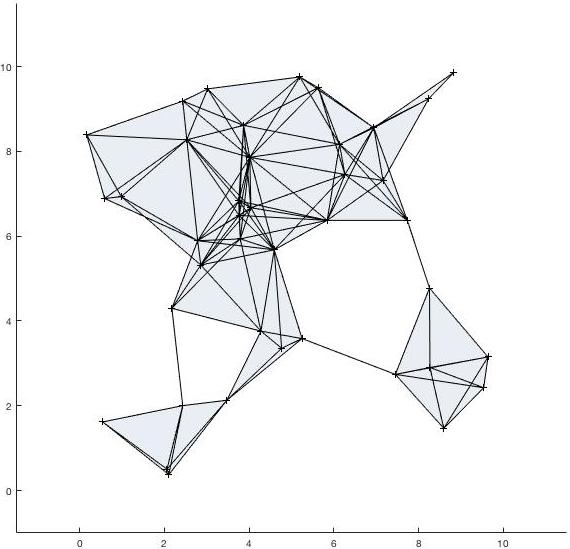}\\
    \includegraphics[width=4.35cm]{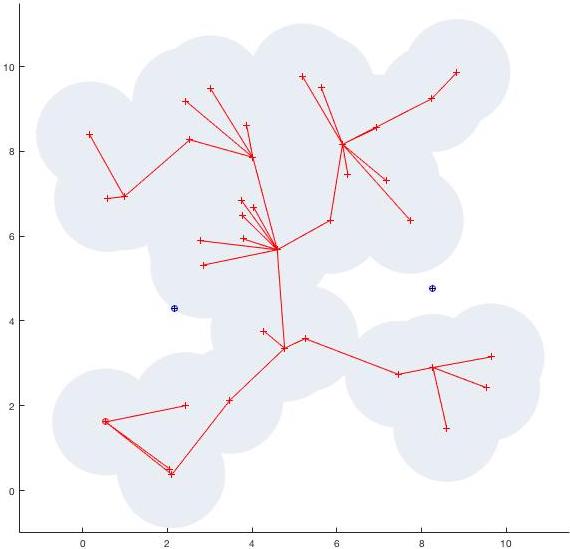}
 \includegraphics[width=4.35cm]{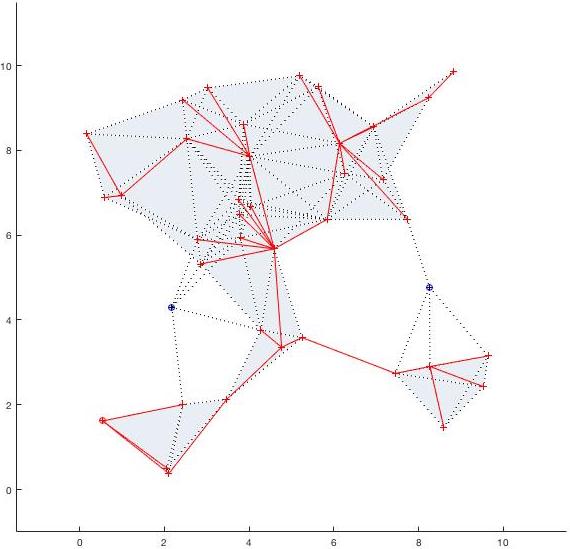}
  \caption{ A coverage hole-free communication tree in a wireless network}
\label{fig_net}
\end{figure}

We can see on the first two figures of Fig.~\ref{fig_net} a wireless
network with two coverage holes and its Vietoris-Rips complex.
The result of our algorithm can be seen in the two following figures with
the tree highlighted in red. There are $2$ vertices (in blue) that
are not in the tree in order to avoid coverage holes. We can verify
that the coverage of the tree is hole-free.
\vspace{-0.1cm}

\vspace{-0.1cm}



\section{Simulation results}
\label{sec_sim}
\subsection{Percentage of rejected vertices}
\begin{figure}[h]
  \centering
    \includegraphics[width=6cm]{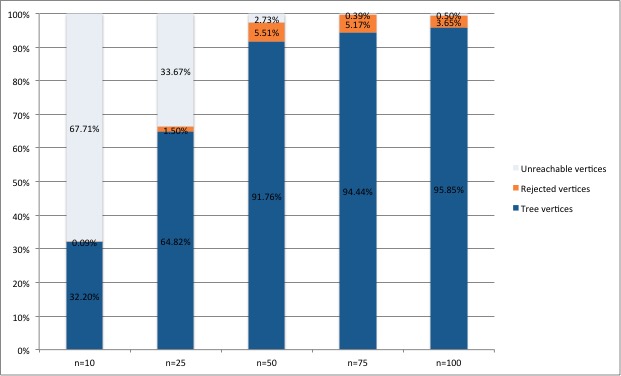}
  \caption{Percentage of unreachable, rejected, and tree vertices.}
\label{fig_rejet}
\end{figure}

For a start, we look at the percentage of vertices that are not in the
final tree at the end of the algorithm. Vertices can be absent from
the final tree for two reasons. First, if vertices are not in the
same connected component as the root vertex, then they are
unreachable. Second, if vertices are in the same 
connected component as the root vertex but create a coverage hole,
they are then rejected by the test on $\beta_1$. We provide in
Fig.~\ref{fig_rejet} a bar chart on which are represented the
percentage of unreachable, rejected, and tree
vertices for different values for the number of
initial vertices $n$, on average on $1000$ simulations for each
scenario. 
The chosen weight metric is the minimum distance, and
the simulation is made on a square of side $a=10$ with a connection
distance of $r=a/4$. 
\vspace{-0.2cm}

We can see that when there is no percolation, very few vertices are
rejected by the algorithm. But when percolation has occurred, that is
when there are almost no unreachable vertices, the percentage of
rejected vertices is below $6\%$ and decreases when the
number of initial vertices grows.

\subsection{Percentage of covered area}
\begin{figure}[h]
  \centering
    \includegraphics[width=6cm]{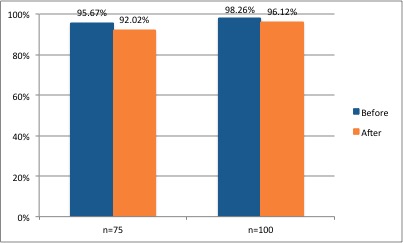}
  \caption{Percentage of covered area before and after the algorithm.}
\label{fig_cover}
\end{figure}

Then we are interested in the loss of coverage that is induced by the
reject of some vertices. To do that we compare the
area covered before the algorithm runs with all vertices, and the area
covered by only the tree vertices. The bar chart in
Fig.~\ref{fig_cover} shows the results for $n=75$ and $n=100$ 
vertices when percolation has occurred there are almost no unreachable
nodes. The configuration is the same as before otherwise.
\vspace{-0.2cm}

We can see that the loss of coverage represents only between $2\%$ and
$3\%$ of the covered area.

\vspace{-0.5cm}
\subsection{Weight metric influence}
\begin{figure}[h]
  \centering
    \includegraphics[width=6cm]{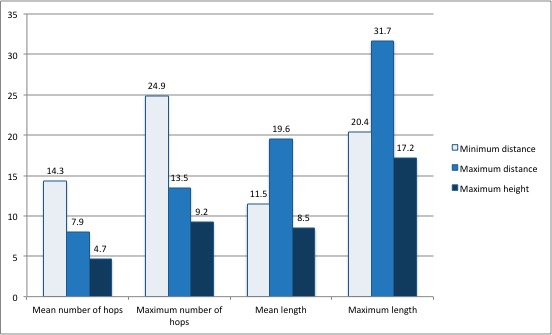}
  \caption{Weight metric influence on the branches of the tree.}
\label{fig_weight}
\end{figure}

Finally, we look into the influence of the chosen weight metric on the
branches on the tree. We compared three weight metric: 
minimum distance, maximum distance, and maximum height. The height of an
edge is defined as the size of the largest simplex it is part of. It
can be interpreted as a redundancy parameter. 
To evaluate the branches, we looked at the mean number of hops, the
maximum number of hops, the mean length and the maximum length.
The results in Fig.~\ref{fig_weight} are given for $n=75$ vertices and the
same parameters as before.
\vspace{-0.2cm}

\vspace{-0.2cm}
We can see that the maximum height minimizes the size of the branches
both in number of hops and in total length. And logically, the minimum
distance maximizes the number of hops, while the maximum
distance maximizes the length of the branches.

\begin{figure}[h]
  \centering
    \includegraphics[width=6cm]{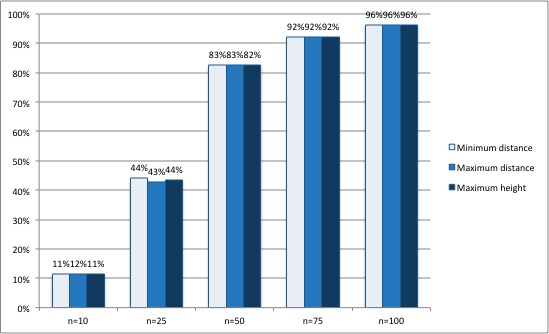}
  \caption{Weight metric influence on the covered surface.}
\label{fig_weightcovered}
\end{figure}

Otherwise the weight metric does not change the size of the covered
area of the final tree as we can see in Fig.~\ref{fig_weightcovered}
whatever the number $n$ of initial vertices. Therefore, since the size
of the covered area is not impacted, the height
seems to be a good metric because long branches are synonyms of
delays and a great number of hops increases the error probability.
\vspace{-0.2cm}

\vspace{-0.3cm}
\section{Communication groups in a network}
\label{sec_forest}
Our coverage hole-free communication tree building algorithm can be
extended to create communication groups in a wireless network. Indeed
a wireless network operator would rather choose several small
communication trees rather than one giant communication tree.
In order to do that with our algorithm, it suffices to limit the
number of hops a branch of the tree can have. Then as long as there are still nodes in the network not in
a communication tree, a new root is randomly chosen among them and a
new tree is created.
\vspace{-0.2cm}
\begin{figure}[H]
  \centering
    \includegraphics[width=4.35cm]{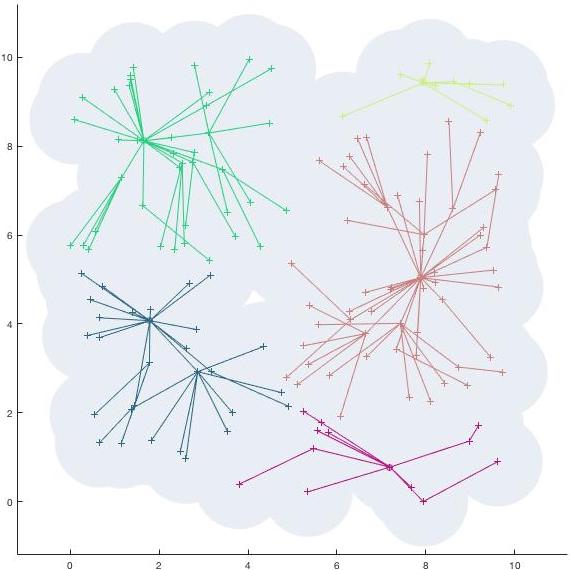}
 \includegraphics[width=4.35cm]{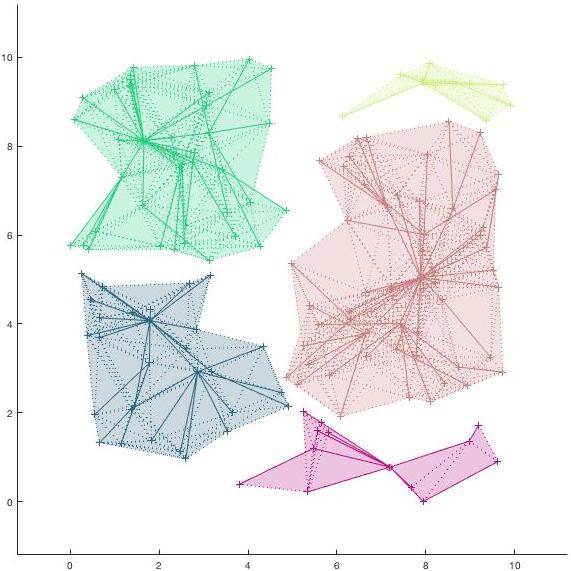}
  \caption{Forest of coverage hole-free communication trees.}
\label{fig_forest}
\end{figure}

At the end, we obtain a forest of small coverage hole-free
communication trees with branches no longer than a given number of
hops as we can see in Fig.~\ref{fig_forest}. The limit number of hops
is set to $3$. Each tree has a different color, their roots, which serve as
communication hubs, are circled. 



\end{document}